\documentclass[fleqn,10pt]{wlscirep}
\usepackage[utf8]{inputenc}
\usepackage[T1]{fontenc}

\usepackage{xcolor}

\usepackage{graphicx}
\usepackage{amssymb} % Need some of symbols including <~
\usepackage{color}
\usepackage{comment}
\usepackage{mcite}
\usepackage{titlesec}
\usepackage{hyperref}
\usepackage{caption}
\usepackage[capitalize]{cleveref}
\usepackage{subcaption}
\usepackage{array}
\usepackage{makecell}

\graphicspath{{Figures/}}

\title{Platinum Black for stray-light mitigation on high-aspect-ratio  micromechanical cantilever}

\author[1,*]{Gautam Venugopalan}
\author[1]{Giorgio Gratta}

\affil[1]{Physics Department, Stanford University, Stanford, California 94305}

\affil[*]{gautamve@stanford.edu}

\begin{abstract}
Microscopic devices are widely used in optomechanical experiments at the cutting-edge of precision experimental physics. Such devices often need to have high electrical conductivity but low reflectivity at optical wavelengths, which can be competing requirements for many commonly available coatings. In this manuscript, we present a technique to electroplate platinum with a highly convoluted surface on a $475\,\mathrm{\mu m } \, \times 500\,\mathrm{\mu m } \, \times 10\,\mathrm{\mu m }$ Silicon/Gold cantilever, preserving its electrical conductivity but reducing its reflectivity in the $0.3 - 1\,\mathrm{\mu m}$ range by a factor of $100$ or greater. The fact that the deposition can be done post-fabrication without damaging delicate structures makes this technique of interest to a potentially large range of experimental applications.
\end{abstract}
\begin{document}

\flushbottom
\maketitle
\thispagestyle{empty}

\section*{Introduction} \label{sec:intro}

Optomechanical experiments employ or directly study the interaction of light with matter and cover a broad range of regimes, from sub-wavelength devices, to kilometer scale, and often aspire to reach or surpass quantum limits~\cite{Aspelmeyer:2012, Gonzalez:2021}. Optical fields are generally required to interact with the mechanical systems of interest in very specific ways, and in many cases the requirement of controlling and limiting stray light arise.

\vspace{1 mm}

\begin{figure}[ht]
    \centering
    \includegraphics[width=0.95\linewidth]{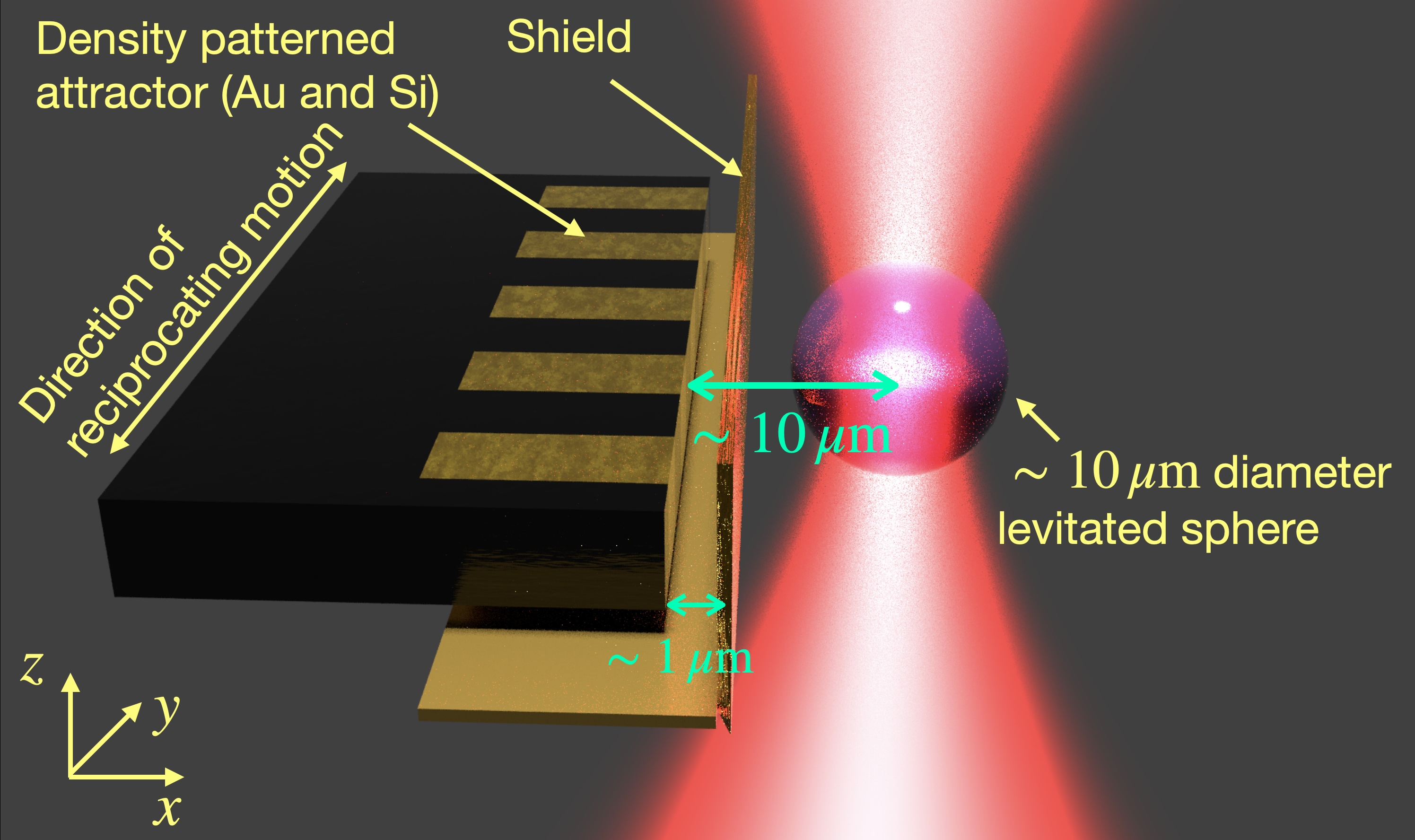}
    \caption{Geometry of experiment (not drawn to scale), with some important components and dimensions annotated, along with the coordinate system. This graphic was rendered with Blender~\protect\cite{Blender:2024} whose ray-tracing does not necessarily reproduce the Mie-scattering regime of the trap.}
    \label{fig:experimentalGeometry}
\end{figure}

\noindent Optical tweezers constitute a class of optomechanical techniques that use tightly-focused laser beams to trap, control, and isolate a test system from its environment~\cite{Ashkin:1970}. The present work is concerned with an experiment levitating $\sim 10\,\mathrm{\mu m}$ diameter silica microspheres (MSs) in vacuum. \Cref{fig:experimentalGeometry} illustrates the geometry of the experiment. Once levitated, the MSs are used as test masses to search for deviations from Newtonian gravity~\cite{Kawasaki:2020} at length scales of a few $\mathrm{\mu m}$. This requires bringing a source-mass, a density-patterned `attractor', close to a levitated MS. As shown in \Cref{fig:experimentalGeometry}, another microfabricated device, a stationary `shield', is placed in between the attractor and MS, for the purposes of shielding the latter from stray electrical fields and gradients sourced by the former. The attractor ($475\,\mathrm{\mu m } \, \times 500\,\mathrm{\mu m } \, \times 10\,\mathrm{\mu m }$, $x \times y \times z$) consists of a cantilever made out of silicon and gold. Various steps involved in the fabrication of the attractor are described in~\cite{Wang:2017}. As the attractor moves with reciprocating motion along $y$, the contrast between the density of silicon, $\rho_{\mathrm Si}$, and that of gold, $\rho_{\mathrm{Au}} \sim 10\cdot\rho_{\mathrm{Si}}$, results in a modulation of the force on the MS, due to gravity or gravity-like interactions.  The modulation is then sensed by the deflection of the light field scattered off the levitated MS in the forward direction, imaged on a segmented photodetector. In the experiment, the surface of the attractor is positioned $\sim 10\, \mathrm{\mu m}$ away from the optical axis of the trapping laser field ($\lambda = 1064\,\mathrm{nm}$), which has a waist of $2w_0\sim 7\,\mathrm{\mu m}$.  This results in a distance of $\sim 5\mathrm{\mu m}$ between the surface of a $10\;\mathrm{\mu m}$ diameter MS and the surface of the attractor.  While great care is taken to ensure good spatial quality of the Gaussian beam in the trapping region, and the shield is also meant to hide the reciprocating attractor from the trapping light, some residual background to the force measurement results from stray reflections of the trapping beam scattering off the attractor and being detected on the photodetector.  Without the shield, it is estimated that $\mathcal{O}(100\,\mathrm{ppm})$ of the (assumed perfectly Gaussian) trap beam would reach the attractor.

\vspace{1 mm}
\noindent A further complication arises from the fact that, as fabricated, the cantilevers have a non-uniform surface topography, presumably due to imperfect Chemical Mechanical Polishing (CMP) after gold was electrochemically grown into trenches etched into the bulk silicon. As a result, even after the attractor is sputter-coated with $\sim 100\,\mathrm{nm}$ of gold (which has an optical depth of $\sim 5\,\mathrm{nm}$ at $\lambda = 1064\,\mathrm{nm}$), the \emph{optical} contrast between the underlying silicon and gold regions is not entirely masked. Consequently, the attractor has coincident density (which sources the signal of the experiment) and reflectivity (which sources backgrounds to the measurement) variations along the density modulated $y$ axis. 

\vspace{1mm}

\noindent The technique of electrodeposition of very rough platinum (``Platinum Black'') has been known for decades to enhance the conductivity of electrodes and appears well suited to provide a light absorbent, electrically conductive, conformal coating that is compatible with the microscopic properties of the attractors.  Extensive investigations have been carried out in order to evaluate the dependence of the growth under various plating conditions~\cite{Feltham:1971}. Here, we present a recipe for performing the plating on a microscopic device with good uniformity and surface quality, and compare its performance to Acktar LithoBlack~\cite{Acktar:2023, Acktar:2024}, a commercially available product for stray-light mitigation.

\section*{Results}

\noindent We used a commercially available plating solution (LabChem LC186807, $3\%\, \mathrm{H_2 PtCl_6}$ , $0.3\%\, \mathrm{Pb(C_2H_3O_2)_2}$). While potentiostats are preferred for precision plating applications, we found that using a voltage source with a large resistance ($100\,\mathrm{k\Omega}$) in series with the plating cell offered sufficient control of the current. In the following sub-sections, we compare the performance and characteristics of Platinum Black coated, and Acktar coated cantilevers (the former coated in-house while the latter obtained commercially). These are also summarized in \Cref{tab:performance}.

\subsection*{Coating thickness}

\begin{figure}[htbp]
    \centering
    \begin{subfigure}[b]{0.45\textwidth}
        \centering
        \includegraphics[width=\textwidth]{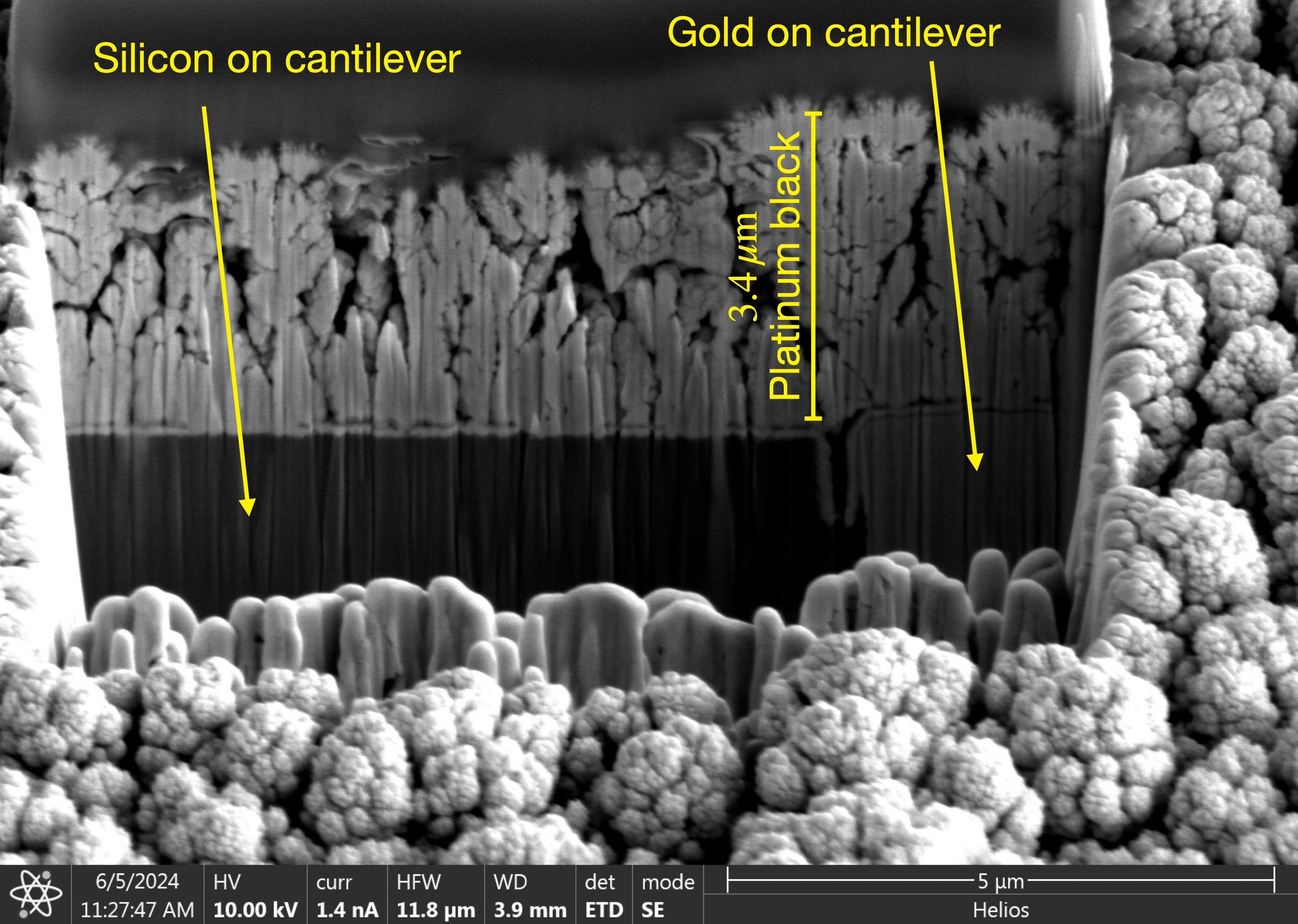} 
        \caption{}
        \label{fig:cross-sec_Pt}
    \end{subfigure}
    \hfill
    \begin{subfigure}[b]{0.45\textwidth}
        \centering
        \includegraphics[width=\textwidth]{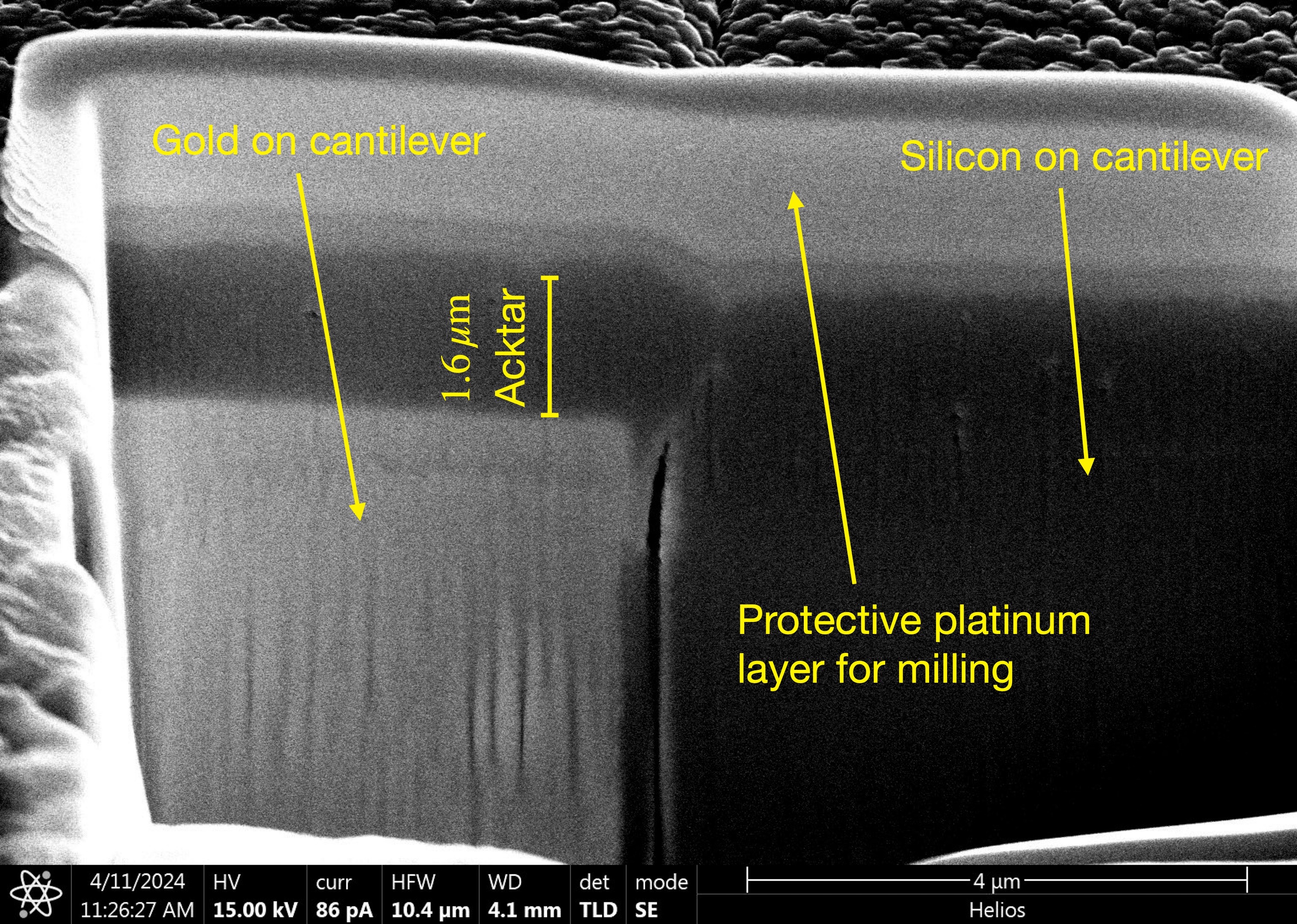} 
        \caption{}
        \label{fig:cross-sec_Acktar}
    \end{subfigure}
    \caption{Milled cross-section showing the thickness of the electrodeposited Platinum Black (\cref{fig:cross-sec_Pt}) and Acktar (\cref{fig:cross-sec_Acktar}) over similar locations on the cantilever. Relevant features on the cantilever, as well as the thickness of the deposited layer are indicated. The contrast has been adjusted to best show the relevant features and the magnification of the two images is slightly different.}
    \label{fig:cross-section}
\end{figure}

A Focused Ion Beam Scanning Electron Microscope (henceforth referred to as FIB SEM, the specific model used was the FEI Helios Nanolab 600i), was used to characterize the thickness of the deposited coating. A protective layer of carbon (for the Platinum Black coating) or platinum (for the Acktar coating) was first deposited and then the cross-section was milled using a $80\,\mathrm{pA}$ beam current. The nature of the protective layer is chosen to maximize contrast with the material under study and has no other function here. A final `cleaning' cross-section cut was performed at $24\,\mathrm{pA}$.  As shown in \Cref{fig:cross-section}, the Platinum Black layer is $\sim 3\,\mathrm{\mu m}$ thick while the Acktar layer is $\sim 1.6\,\mathrm{\mu m}$ thick, both in line with the requirement. In the foreground of~\Cref{fig:cross-sec_Pt}, the very convoluted, high surface area of the deposited platinum is evident, presumably leading to the low reflectivity at optical wavelengths. For comparison, the Acktar layer's morphology is shown in \Cref{fig:cross-sec_Acktar}, and appears smoother.

\subsection*{Coating uniformity}

\begin{figure}[htbp]
    \centering
  
    \begin{subfigure}[b]{0.3\textwidth}
        \centering
        \includegraphics[width=\textwidth]{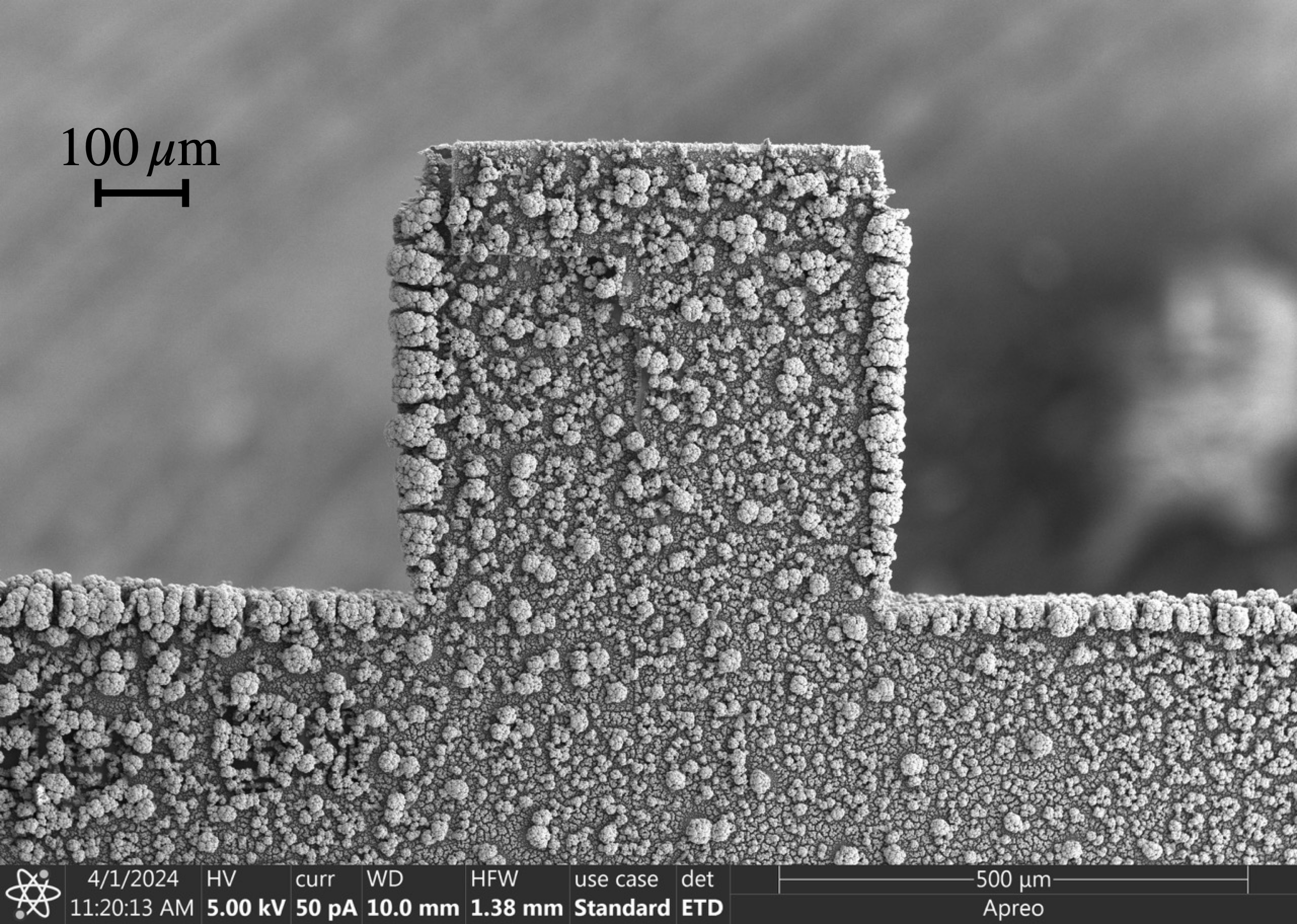} 
        \caption{}
        \label{fig:veryRoughEdges}
    \end{subfigure}
    \hfill
    % Second subfigure
    \begin{subfigure}[b]{0.3\textwidth}
        \centering
        \includegraphics[width=\textwidth]{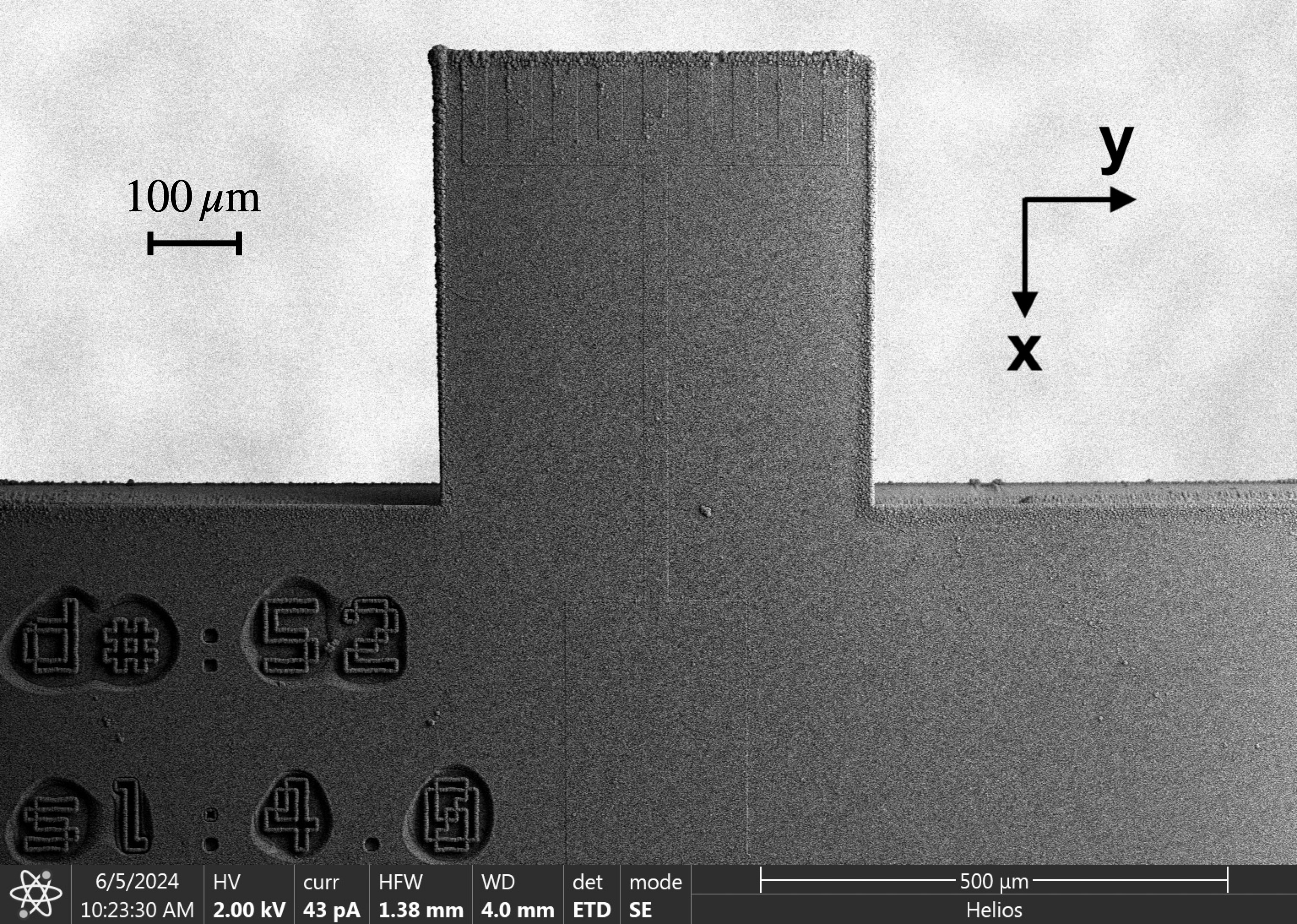} 
        \caption{}
        \label{fig:roughEdges}
    \end{subfigure}
    \hfill
    % Third subfigure
    \begin{subfigure}[b]{0.3\textwidth}
        \centering
        \includegraphics[width=\textwidth]{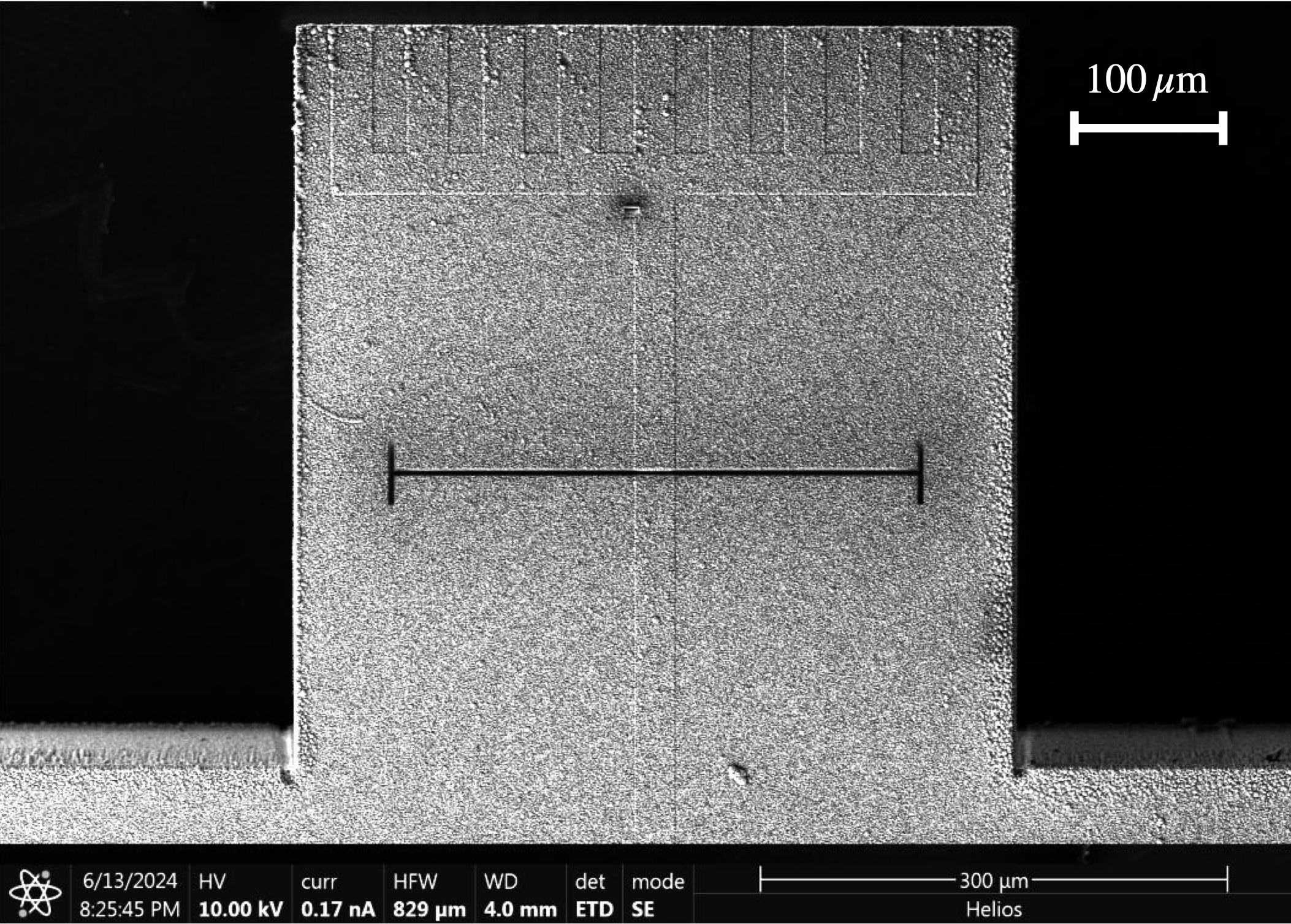} 
        \caption{}
        \label{fig:smoothEdges}
    \end{subfigure}

    \caption{Electron micrographs of electrodeposited platinum black under various conditions, and after different processing steps. Note that the contrast has been adjusted to better show important features, and the magnification varies slightly between images. Scale bars have been added for easier visibility. See text for full description.}
    \label{fig:edgeEffects}
\end{figure}

A known~\cite{Ilic:2020} issue with electrodeposition on high-aspect-ratio geometries is that the growth tends to happen at higher rates around corners where the electric field is higher. This was found to be the case during initial plating tests, as shown in \Cref{fig:veryRoughEdges}. In order to mitigate such non-uniform growth, the plating cell was immersed in a low-power ultrasonic bath (Magnasonic MGUC500, $42\,\mathrm{kHz}$ ultrasound frequency, $\sim 25\,\mathrm{W}$ delivered to a 600~ml volume). While ultrasonication was previously used~\cite{Arcot:2010} to improve the uniformity of the deposited layer, using a low-power bath and restricting the duration of sonication to a few minutes were necessary precautionary measures to avoid damaging the $10\,\mathrm{\mu m}$ thick cantilevers. The deposition rate is known to be very sensitive to the specific concentration of reagents in the plating solution used~\cite{Feltham:1971}. Empirically, the most uniform growth was achieved with a pulsed current, using a square waveform with $50\%$ duty cycle, $1\,\mathrm{mA}$ amplitude, corresponding to an areal current density of $\sim 200\,\mathrm{mA/cm^2}$ for the cantilever geometry. The plating proceeded for $\sim 300\,\mathrm{s}$, suggesting a deposition rate of $\sim 10\,\mathrm{nm/s}$, assumed uniform. 

\noindent A comparison between two different plating protocols is shown in~\Cref{fig:veryRoughEdges,fig:roughEdges}. For the device in~\Cref{fig:veryRoughEdges}, a DC plating current was used with no ultrasonication, while \Cref{fig:roughEdges} shows the result when using a combination of pulsed plating current and having the cell immersed in an ultrasonic bath during the plating.

\vspace{1mm}
\noindent Even after the above improvements to the plating setup, some slight excess growth in the corners of the cantilever persisted, 
while the experiment requires the geometry of the cantilever be rectangular to better than $1\,\mathrm{\mu m}$ in the $x-$coordinate over the $500\,\mathrm{\mu m}$ width ($y-$coordinate) of the device (see \Cref{fig:roughEdges} for coordinate system). To realize this uniformity, the FIB SEM was used to mill off excess growth. The relatively light milling species (Gallium) and the fact that a large volume of material had to be removed ($\sim 3\,\mathrm{\mu m} \times 500\,\mathrm{\mu m} \times 15\,\mathrm{\mu m} = 22500\,\mathrm{\mu m ^3}$) meant that this process took several hours, even at the maximum milling current of $65\,\mathrm{nA}$. Cleaning cuts had to be done at the end of the process at a lower milling current of $2.5\,\mathrm{nA}$ to reduce curtaining and achieve a smoother finish. A heavier milling species (e.g. Xe, which is available in certain Plasma Focused Ion Beam (PFIB) setups such as the FEI Helios Hydra) would considerably speed up the process. The uniformity and surface finish of the cantilever after ion-beam milling can be seen in~\Cref{fig:smoothEdges}. The additional feature seen halfway along the cantilever in~\Cref{fig:smoothEdges,fig:opticalMicroscopeCoated} (absent in~\Cref{fig:roughEdges,fig:opticalMicroscopeUncoated}) is a $335\, \mathrm{\mu m}$ long fiducial mark milled into the cantilever for later optical metrology.

\subsection*{Optical reflectivity}

\vspace{1 mm}

\begin{figure}[ht]
    \centering
    \includegraphics[width=0.95\linewidth]{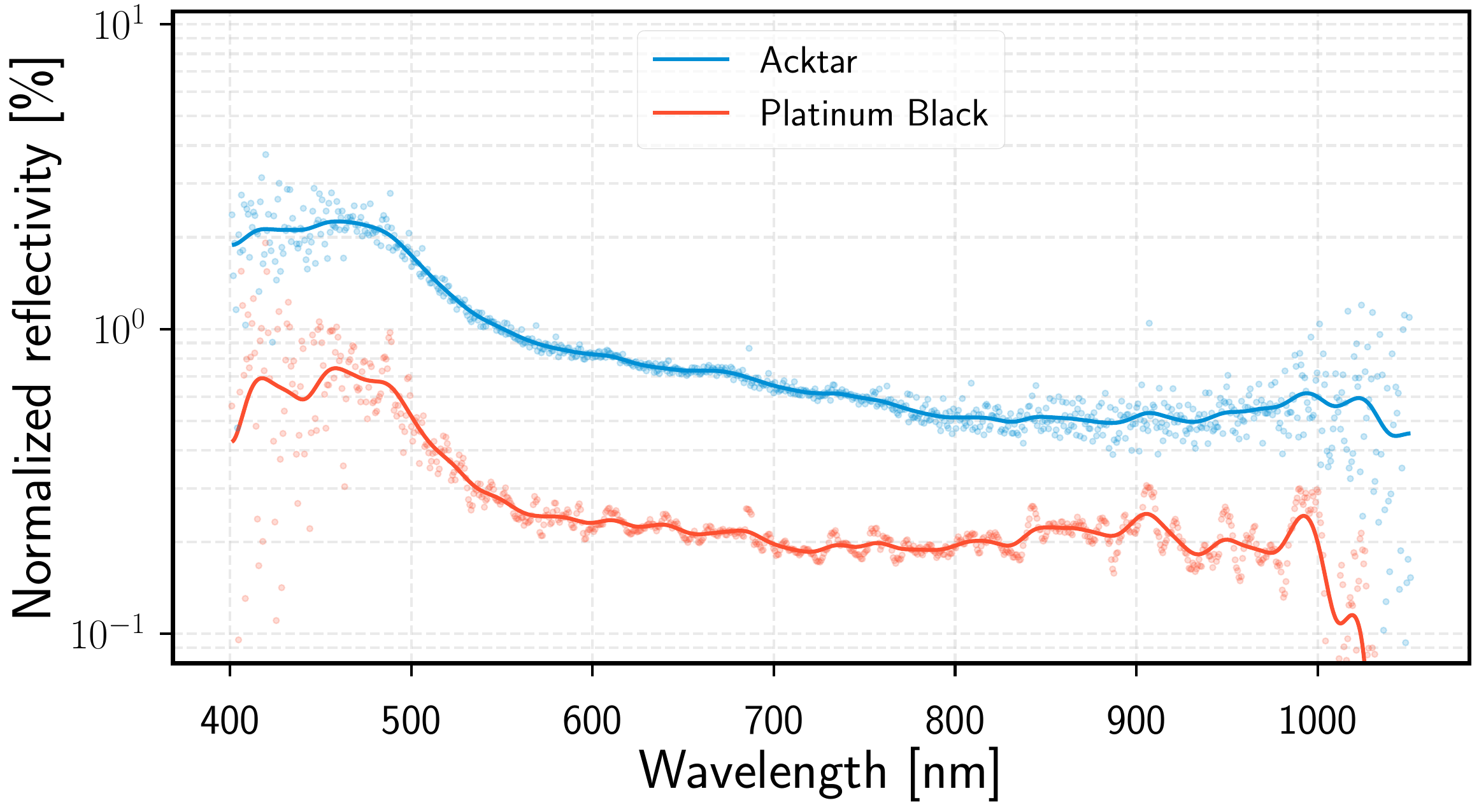}
    \caption{Reflectometry data for Platinum Black- and Acktar-coated cantilevers, normalized to that of an gold-coated cantilever. Measured datapoints are shown as light circles, while the solid lines represent the same information, smoothed using a Gaussian kernel for easier visual interpretation.}
    \label{fig:directReflComparison}
\end{figure}

Reflectometry measurements were done using a Filmetrics F40, the results of which are shown in~\Cref{fig:directReflComparison}. The reflectivity of each coating is shown normalized to that of a device coated with 100~nm of sputtered gold.  The devices with additional Actkar or Platinum Black coatings exhibit $\gtrsim 100 \, \times $ lower broadband reflectivity in the $0.4 - 1.1\,\mathrm{\mu m}$ range.  Platinum Black has slightly better performance than Acktar, though this may be due to the latter being slightly thinner (measured to be $\sim 1.6\,\mathrm{\mu m}$) by FIB cross-sectioning, versus the $\sim 3\,\mathrm{\mu m}$ thick platinum black.

\vspace{1 mm}

\begin{figure}[htbp]
    \centering
    \begin{subfigure}[b]{0.45\textwidth}
        \centering
        \includegraphics[width=\textwidth]{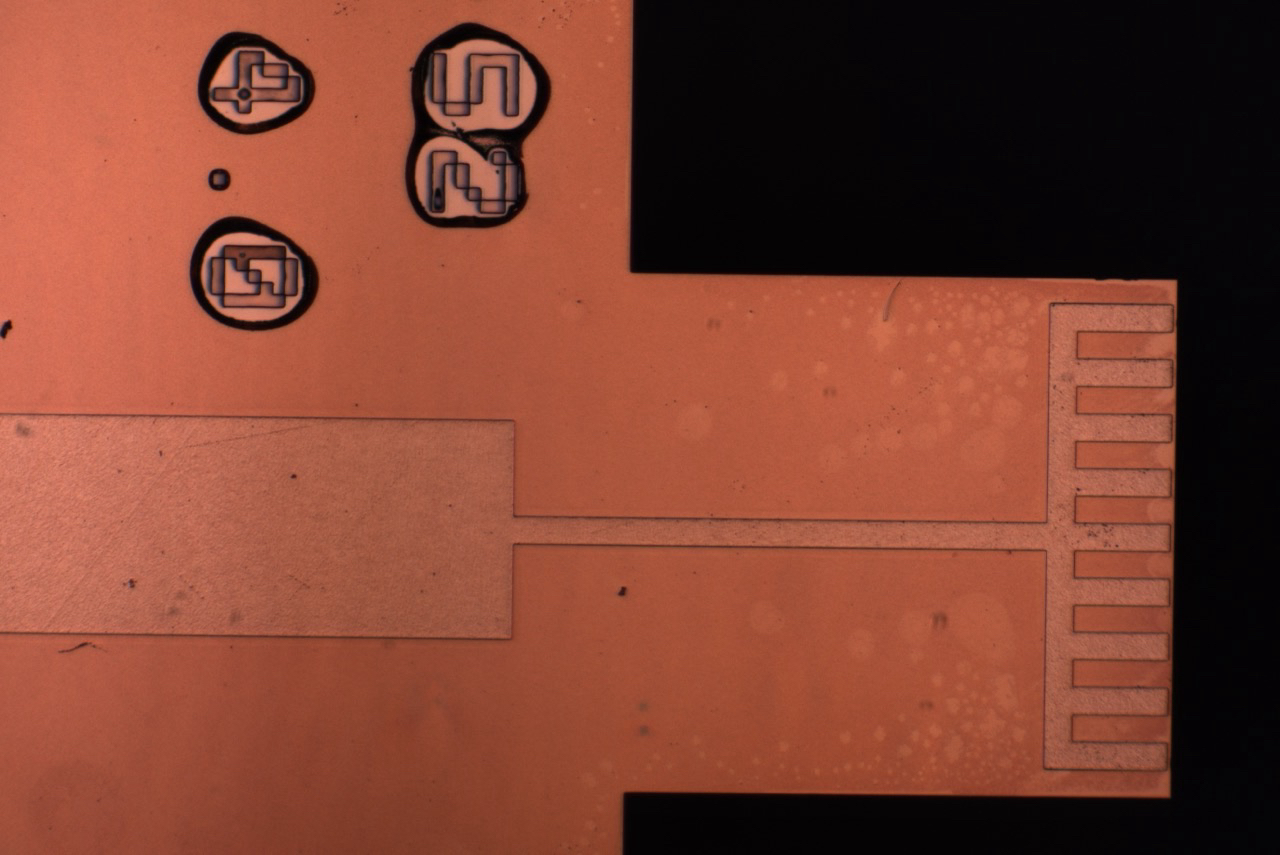} 
        \caption{}
        \label{fig:opticalMicroscopeUncoated}
    \end{subfigure}
    \hfill
    % Third subfigure
    \begin{subfigure}[b]{0.45\textwidth}
        \centering
        \includegraphics[width=\textwidth]{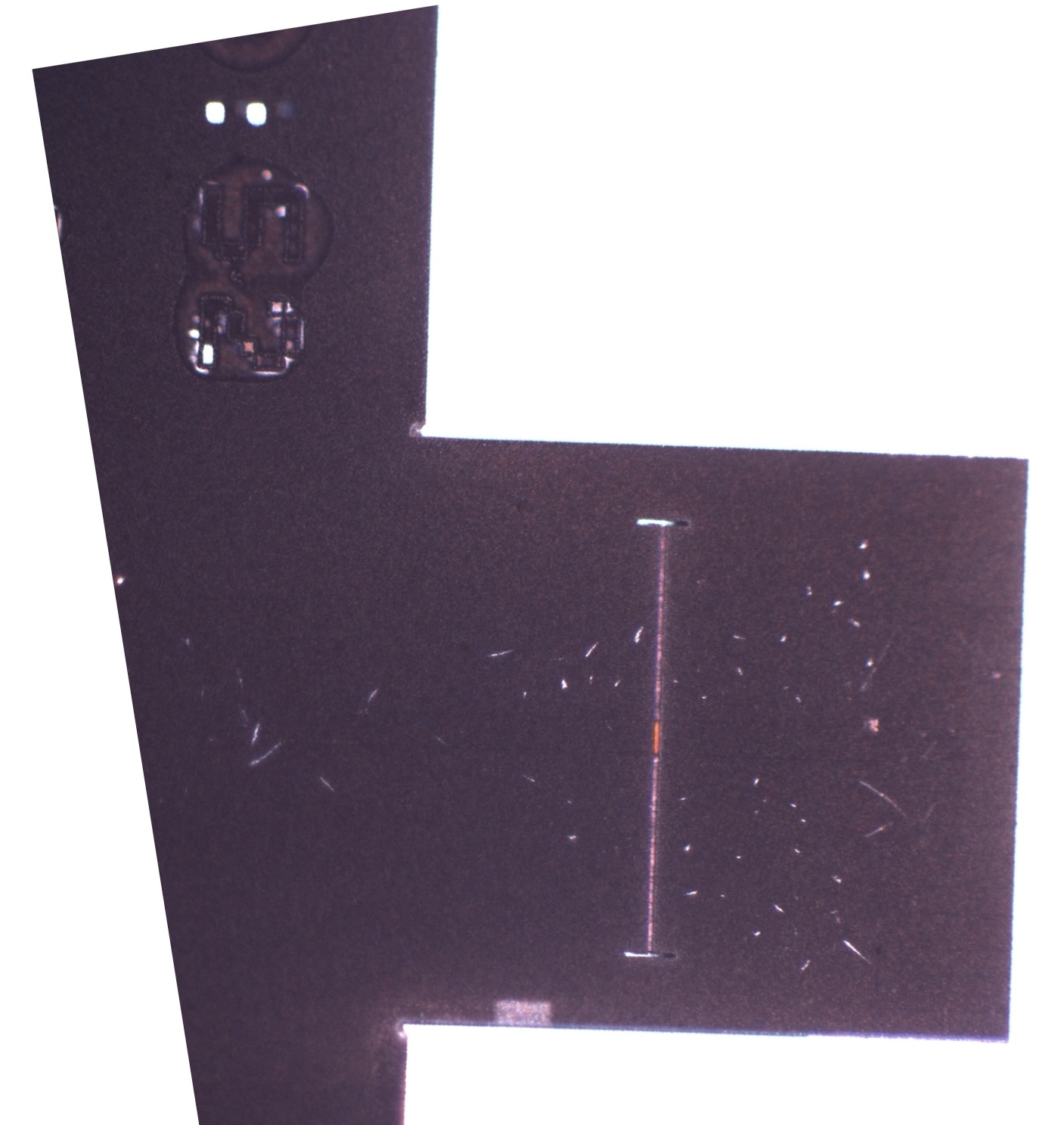} 
        \caption{}
        \label{fig:opticalMicroscopeCoated}
    \end{subfigure}

    \caption{Characterization of coated cantilever using a $10\,\times$ microscope objective paired with a CMOS camera. See text for full description.}
    \label{fig:ex-situ-metrology}
\end{figure}

\noindent Additionally, images of the cantilever before and after coating with Platinum Black were captured using an optical microscope and CMOS camera, shown in~\Cref{fig:opticalMicroscopeUncoated,fig:opticalMicroscopeCoated} respectively. The image of the coated cantilever was acquired with an exposure time $\sim 80\times$ longer than that for the uncoated device. The brightness of the image qualitatively illustrates the lower reflectivity of the coated device.

\subsection*{Electrical conductivity}

\vspace{1mm}

Because of the possibility of electrostatic background to the measurement, it is desirable for the light-absorbent coating to preserve the surface electrical conductivity of the original sputtered gold.  Both attractor and shield have their gold surface electrically accessible to set their potentials (in particular, the potential difference between them). These connections allow for an in-situ measurement of electrical continuity between the two microscopic devices, when they are brought into mechanical contact with the help of nanopositioning stages. For reference, the resistance between the uncoated cantilever and the shield was measured to be $\sim 20\Omega$, which is representative of the sheet resistance of sputtered gold. While the exact area of contact is unknown, the attractor-shield contact is measured (using a Keithley 2100 multimeter) to have a resistance of $\sim 10\,\Omega$ in the case of the Platinum Black coated attractor, and $>100\,\mathrm{M\Omega}$ in the case of the Acktar coated attractor. The sheet resistivity of the Acktar LithoBlack coating is specified~\cite{Acktar:2024} to be $\leq 2\,\mathrm{M\Omega/\square}$, suggesting an aspect ratio of the contact area $\geq 50$, which is not unreasonable for the geometry. The same aspect ratio would suggest a sheet resistivity for the electrochemically deposited Platinum Black to be $\sim 0.2\Omega/\square$.

\begin{table}[h!]
    \centering
    \begin{tabular}{>{\centering\arraybackslash}m{0.2\textwidth} >{\centering\arraybackslash}m{0.3\textwidth} >{\centering\arraybackslash}m{0.2\textwidth} >{\centering\arraybackslash}m{0.2\textwidth} }
    \hline
    \textbf{Property} & \textbf{Requirement} & \textbf{Platinum Black} & \textbf{Acktar} \\
    \hline
    \hline
    Coating thickness & $\leq{5\,\mathrm{\mu m}}$ on each side.  & $\sim 3.4\,\mathrm{\mu m}$ & $\sim 1.6\,\mathrm{\mu m}$ \\
    \hline
    Coating conformality & Maximally conformal, with deviations $\leq{1\,\mathrm{\mu m}}$ in any of the 3 cartesian axes. & After optimized plating setup and ion-beam milling, conformal & Conformal \\
    \hline
    Optical reflectivity & Best effort, target $\leq 1\%$ that of Au-coated cantilever at $\lambda=1064\,\mathrm{nm}$.  & $\leq 0.5\%$ & $\sim 1\%$\\
    \hline
    Electrical conductivity & Best effort, highest possible.  & Conductive, estimated sheet resistance $\sim 0.2\,\Omega/\square$ & Insulating, estimated sheet resistance $\sim 2\,\mathrm{M\Omega/\square}$ \\
    \hline
    Yield without damage & Best effort, highest possible for devices already diced from wafer. &  With optimized plating setup, $\sim 80\%$ ($n=5$ samples) & $\sim 10\%$ ($n=10$ samples) \\
    \hline
    \end{tabular}
    \caption{Summary of requirements and results obtained with two candidate coating options.}
    \label{tab:performance}
    \end{table}

\section*{Discussion}

We have demonstrated a method to coat microscale devices with Platinum Black to absorb visible and near IR light. In the process, the optical reflectivity relative to gold was reduced by $\sim 100\,\times$, while the surface remained electrically conductive and the device was not mechanically damaged during the coating process. 

This technique may be of interest to other applications in which stray-light control from micro- or nano-mechanical devices is of importance, and other coating options are not viable. 

\section*{Competing interests}

The authors declare no competing interests.

\section*{Data availability}
Electron micrographs, optical images, and reflectivity data presented in this study are available from the corresponding author on reasonable request.

\bibliography{main}

\section*{Acknowledgements}

The authors would like to thank E.J.~Chichilnisky, Lim Ye-Lim, and Praful Vasireddy for helpful discussions about the coating process and for supplying initial batches of the plating solution. Paul Wallace and Juliet Jamtgaard provided valuable input in early ion-milling trials.

\vspace{1mm}

\noindent The authors would also like to thank Qidong Wang for pointing out prior work that performed plating in an ultrasonic bath for improved uniformity of the growth, and Eric Hoppe for helpful discussions while putting together the initial plating setup. The batch of devices used in this work was fabricated by Chas Blakemore and Alex Rider. Clarke Hardy, Jacqueline Huang, Kenneth Kohn, Lorenzo Magrini, Albert Nazeeri	, Zhengruilong Wang, and Yuqi Zhu contributed many insights in framing the problem.  
\vspace{1mm}

\noindent This work was supported by NSF Grant 2406999 and ONR Grant N000142312600. Part of the work was performed at the Stanford Nano Shared Facilities (SNSF) which is supported by the NSF under award ECCS-2026822

\section*{Author contributions statement}

G.V. and G.G. conceived the idea of depositing Platinum Black on the electrodes. G.V. constructed the plating setup and performed the plating, ion-beam milling, and reflectivity characterization. All authors reviewed the manuscript.

\end{document}